\documentclass[12pt]{iopart}
\usepackage{graphicx}
\usepackage{dsfont}
\usepackage{cite}

\begin{document}
\title[Local Hamiltonians for one-dimensional critical models]{Local Hamiltonians for one-dimensional critical models}

\author{Dillip K. Nandy$^1$,  N. S. Srivatsa$^2$ and Anne E. B. Nielsen$^{2,}$\footnote{On leave from: Department of Physics and Astronomy,
Aarhus University, Ny Munkegade 120, Denmark, 8000 Aarhus C}}
\address{$^1$ Department of Physics and Astronomy, Aarhus University, Ny Munkegade 120, DK-8000 Aarhus C}
\vspace{10pt}
\address{$^2$ Max-Planck-Institut f\"{u}r Physik komplexer Systeme, D-01187 Dresden, Germany}

\begin{abstract}
Conformal field theory has turned out to be a powerful tool to derive interesting lattice models with analytical ground states.
Here, we investigate a class of critical, one-dimensional lattice models of fermions and hardcore bosons related to the Laughlin states.
The Hamiltonians of the exact models involve interactions over long distances that are difficult to realize experimentally. This motivates us
to study the properties of models with the same type of interactions, but now only between nearest and possibly next-nearest neighbor sites.
Based on computations of wavefunction overlaps, entanglement entropies, and two-site correlation functions for systems of up to 32 sites, we
find that the ground state is close to the ground state of the exact model. There is also a high overlap per site between the lowest excited
states for the local and the exact models, although the energies of the low-lying excited states are modified to some extent for the system
sizes considered. We also briefly discuss possibilities for realizing the local models in ultracold atoms in optical lattices.

\vspace{2pc}
\noindent{\textbf{Keywords}}: Spin chain models, conformal field theory, exactly solvable models, entanglement entropies, correlation functions,
ultracold atoms in optical lattices
\end{abstract}



\section{Introduction}
The physics that can be realized in strongly-correlated quantum many-body systems is rich and far from being fully understood.
The systems are challenging to investigate, because the resources needed to do numerical studies of the systems are often at
the borders of or beyond what can be done. Analytical models therefore play an important role in getting insight into the physics of the systems.

Conformal field theory has turned out to be an interesting tool to study quantum many-body systems. The idea to interpret certain conformal field theory correlators as wave functions dates back to the work by Moore and Read \cite{read}, and conformal field theory also plays an interesting role in the Haldane-Shastry model \cite{spincft}. A particular technique to construct parent Hamiltonians for different quantum many-body models on lattices was proposed in \cite{Nielsen}. In these models, the ground state, and sometimes also excited states, are conformal field theory correlators. The Hamiltonians typically consist of terms that are few-body and nonlocal. The method has, for instance, been used to obtain 2D lattice models with analytical ground states that are lattice versions of the Laughlin state at filling factor $1/q$, where $q=2,3,\ldots$ \cite{TuNJP}. These states are topologically ordered and can host Abelian anyons.

One can also obtain 1D versions of the lattice Laughlin models by doing the same derivation in one dimension \cite{TuNJP}.
The resulting models are critical and have interactions that decay as the inverse of the square of the distance ($1/r^2$). The model
for $q=2$ turns out to coincide with the famous Haldane-Shastry model \cite{Haldane,Shastry,Talstra}. The Haldane-Shastry model
is interesting, since all the excited states of the model can be found analytically, and since the model allows for semionic
excitations. It is also worth mentioning that the low-energy properties of the Haldane-Shastry model and the nearest-neighbor antiferromagnetic $XXX$ Heisenberg model fall in the same universality class and are described by the SU$(2)_1$ Wess-Zumino-Witten model with central charge $c=1$ \cite{FDM}. It is likely that the excited states of the one-dimensional models for other values of $q$ are also interesting and related to the anyons that can be obtained in the 2D versions of the models. One-dimensional quantum many-body models with nonlocal interactions of the
kind $1/r^2$, as well as truncated versions of them, have gained considerable interest over the
years \cite{Calogero, Sutherland, Haldane, Shastry, Talstra,FDM, Wang, Cirac, Nielsen, TuNJP, Thomale, Pittman, Tummuru}, exploring
interesting aspects of integrable systems. A common feature of quantum $1/r^2$ systems is that the ground state can be exactly
represented by a Jastrow-type wavefunction \cite{Calogero, Sutherland, Haldane, Shastry, Talstra,FDM, Wang, Cirac, Nielsen, TuNJP,
Thomale, Pittman, Tummuru}, namely a product of two-body functions.

One purpose of constructing analytical models of quantum many-body systems with interesting properties is to use them as a guide to find
ways to implement the physics experimentally. In many cases, it may not be possible to directly implement the analytical model itself, but
with some suitable modifications one may be able to simulate a model displaying the same physics. Ultracold atoms in optical lattices provide
an interesting and flexible setup for simulating quantum models, and there is a rapid development in the field \cite{Lewenstein,Gross}.
As far as the models derived from conformal field theory are concerned, a challenging aspect for experimental implementation is that the
models involve nonlocal interactions. This poses the question, whether the same physics can be realized in a model with only local interactions, and whether one can use the form of the exact model to predict a suitable choice of local interactions. In this article, we investigate this question for the family of 1D models related to the Laughlin states. We do the investigation with exact diagonalization, and we find that it is possible to obtain local Hamiltonians, whose low-energy eigenstates have high overlap per site with the eigenstates of the exact Hamiltonians. For the special case $q=2$, the truncation transforms the Haldane-Shastry model into the nearest-neighbor antiferromagnetic $XXX$ Heisenberg model, and as mentioned above, these models are known to fall in the same universality class.

Investigations of whether a nonlocal 1D Hamiltonian derived from conformal field theory can be replaced by a local Hamiltonian have been done for other systems previously with affirmative conclusions \cite{Cirac, Nielsen, TuNPB, Glasser}. These investigations have focused mainly on models with SU(2) or SU($n$) symmetry or spin models with a symmetry when flipping all the spins and have looked only at ground state properties and mainly ground state overlaps. In the present article, we do a more thorough investigation for a family of models without these
symmetries (the model with $q=2$ has SU(2) symmetry, but the others do not). The presence of symmetries simplifies the study, since the symmetry reduces the number of possible local interactions, and this makes it interesting to ask the question also for models with less symmetry. In addition to ground state overlap, we compute entanglement entropy and correlation functions for the ground states of the local models, which are important
quantities for determining the physics of the models. The low-energy excited states are, however, also crucial when considering dynamics or systems at low, but nonzero, temperature. We therefore also compute the low-lying part of the energy spectra and overlaps for the low-lying excited states. Finally, we briefly discuss possibilities for implementing the local models in ultracold atoms in optical lattices. Implementing the models would
allow the power law decay of the correlation functions and the logarithmic growth of entanglement entropy with subsystem size
to be measured \cite{Altman,Sherson,Islam}. In addition, one could investigate the low-lying excited states, such as the semionic excitations for $q=2$. 

The results presented in this article suggest that the interesting physics of the exact models can be realized in models with local interactions.
The studies are limited to systems of up to $32$ lattice sites, as we are using exact diagonalization to compute the eigenstates of the local models.
Implementing the models experimentally would allow even larger systems to be considered and could hence provide further insight into how well the
local models reproduce the physics of the exact models.

The paper is organized as follows. In Sec.\ \ref{sec:model}, we introduce the family of exact Hamiltonians and their analytical ground states.
We show that the interaction strengths decay with distance, and we propose different local models with nearest-neighbor (NN) and possibly
next nearest-neighbor (NNN) couplings. We also briefly discuss possibilities for realizing the local models in ultracold atoms in optical
lattices. In Sec.\ \ref{sec:gsprop}, we quantify how well the local models capture the physics of the nonlocal models by computing properties
of the ground states, including overlap, entanglement entropy, and correlation functions. In Sec.\ \ref{sec:esprop}, we compute the lowest
energies of the models and the overlap between excited states of the local and the exact models. Section \ref{sec:conclusion} concludes the paper.

\section{Model Hamiltonians}\label{sec:model}
\subsection{Exact model}
Our starting point is a family of exact 1D models with nonlocal interactions that have been constructed using tools from a
conformal field theory in \cite{TuNJP}. The models are defined on a 1D lattice with periodic boundary conditions, and it is convenient to
think of the $N$ lattice sites as sitting on a unit circle in the complex plane with the $j$th site at $z_j = e^{i2\pi j/N}$. Each site can
be either empty or occupied by one particle, and $n_j$ denotes the number of particles on site $j$. The members of the family are labeled
by a positive integer $q$, and for $q$ odd (even), the particles are fermions (hardcore bosons).

The Hamiltonian takes the form
\begin{eqnarray}
 H_{\textrm{1D}} = \sum_{i \ne j}
 \left[
 (q-2)w_{ij} - w_{ij}^2
 \right]
 d^{\dagger}_i d_j
 - \frac{1}{2}(q^2 - q) \sum_{i \ne j} w_{ij}^2 n_i n_j. \label{Ham}
\end{eqnarray}
Here, $d^{\dagger}_i d_j$ is the particle hopping operator from site $j$ to site $i$, the operator $n_in_j$ denotes
the density-density interaction terms between sites $i$ and $j$, with $n_i=d_{i}^{\dagger}d_{i}$, and $w_{ij} = \frac{z_i + z_j}{z_i - z_j}$. Explicitly,
\begin{eqnarray}
d_j &=&
\left(\begin{array}{cc}
1 & 0 \\
0 & 1
\end{array}\right)_1 \otimes
\left(\begin{array}{cc}
1 & 0 \\
0 & 1
\end{array}\right)_2 \otimes
\ldots \otimes
\left(\begin{array}{cc}
1 & 0 \\
0 & 1
\end{array}\right)_{j-1} \otimes
\left(\begin{array}{cc}
0 & 0 \\
1 & 0
\end{array}\right)_j \nonumber \\
&\otimes&
\left(\begin{array}{cc}
(-1)^q & 0 \\
0 & 1
\end{array}\right)_{j+1} \otimes
\ldots \otimes
\left(\begin{array}{cc}
(-1)^q & 0 \\
0 & 1
\end{array}\right)_{N}.
\end{eqnarray}
For the case of fermions (q odd), the annihilation operator ($d_i$) and the creation operator ($d^{\dagger}_j$) satisfy the usual anti-commutation relation
\begin{eqnarray}
 \{d_i, d^{\dagger}_j \} = \delta_{ij}, ~~ d^2_i = {d^{\dagger}_i}^2 = 0. \label{anti}
\end{eqnarray}
For the case of hard-core bosons (q even), the corresponding relation is given as follow. At different sites the creation and annihilation operators commute as usual for bosons:
\begin{eqnarray}
 \left[d_i, d^{\dagger}_j \right] = \left[d_i, d_j \right] = \left[d^{\dagger}_i, d^{\dagger}_j \right] = 0, ~~~~~ i \ne j, \label{commu}
\end{eqnarray}
and on the same site these operators satisfy anticommutation relations as for fermions in (\ref{anti}). The Hamiltonian conserves the number of particles, and we shall assume throughout that the number of particles in the system
is fixed to $N/q$.

It has been shown analytically in \cite{TuNJP} that the state
\begin{eqnarray}\label{Ana_eqn}
|\Psi_{\textrm{Exact}}\rangle=
\sum_{n_1,n_2,\ldots,n_N}
\Psi_{\textrm{Exact}}(n_1, n_2, \ldots, n_N)
|n_1,n_2,\ldots,n_N\rangle
\end{eqnarray}
with
\begin{eqnarray}
\Psi_{\textrm{Exact}}(n_1, n_2, \ldots, n_N)
\propto \delta_n \chi_n \prod_{i<j}
(z_i - z_j)^{qn_in_j-n_i-n_j},
\end{eqnarray}
is an eigenstate of the Hamiltonian (\ref{Ham}) with energy
\begin{eqnarray}
E_0 = - \frac{(q-1)}{6q} N[3N + (q-8)],
\end{eqnarray}
and numerical investigations for small systems show that this state is the unique ground state in the subspace with $N/q$ particles.
The factor $\delta_n$ is defined by
\begin{eqnarray}
\delta_n &=&
\left\{\begin{array}{ll}
\displaystyle
1 & \mbox{for } \sum_i n_i= N/q
\\ [2ex]
\displaystyle
0 & \mbox{otherwise,}
\end{array}\right.
\end{eqnarray}
and it fixes the number of particle in the state to $N/q$. The factor $\chi_n = (-1)^{\sum_j(j-1)n_j}$ is a sign factor. A subset of the excited
states of the model can also be found analytically as discussed in \cite{TuNJP}.

We comment that the Hamiltonian for $q=2$ is closely related to the Haldane-Shastry spin Hamiltonian \cite{Haldane,Shastry}. This can be seen by introducing the transformation
\begin{equation}
S^{+}_{i}=d_{i}^{\dagger}\\
S^{-}_{i}=d_{i}\\
S^{z}_{i}=d^{\dagger}_{i}d_{i}-1/2, \label{trans}\\
\end{equation}
where $S^{\pm}_{i}=S^{x}_{i}\pm \rmi S^{y}_i$, and $\vec{S}_i=(S^{x}_i,S^{y}_i,S^{z}_i)$ is the spin operator acting on a two-level system. Inserting this into (\ref{Ham}) and utilizing that the number of particles is fixed to $N/2$, we arrive at the Hamiltonian
\begin{equation}
H=\sum_{i\neq j}\frac{\vec{S_i}\cdot\vec{S_j}}{\tan^2(\frac{\pi(i-j)}{N})} + \textrm{constant}.
\end{equation}
This is the Haldane-Shastry Hamiltonian except for an additive and a multiplicative constant.

Monte Carlo simulations for systems with a few hundred sites have already shown that the exact model is critical with a ground state
entanglement entropy that grows logarithmically with the size of the subsystem and ground state two-site correlation functions that
decay as a power law \cite{TuNJP}. In the following, we concentrate on the cases $q=3$, $2$, and $4$. We do so, because we want to study
both the fermionic and the hardcore bosonic case, and because the $q=2$ model has an SU(2) invariant ground state, while
the $q=4$ and $q=3$ models do not.

Let us investigate the behavior of the coefficients $C_1 = (q-2)w_{ij} - w_{ij}^2$ of the hopping terms and $C_2 = -\frac{1}{2}(q^2 - q)w_{ij}^2$
of the density-density interaction terms of the exact Hamiltonian (\ref{Ham}) with respect to distance. First we note that
\begin{equation}
w_{jk}=\frac{z_j+z_k}{z_j-z_k}=\frac{-i}{\tan[\frac{\pi}{N}(j-k)]}.
\end{equation}
We have
\begin{equation}
w_{jk}\approx \frac{-iN}{\pi(j-k)} \qquad \textrm{for} \qquad \pi|j-k|/N\ll 1.
\end{equation}
In this limit, the terms proportional to $w_{ij}^2$ are hence much larger than the term proportional to $w_{ij}$. It follows that $C_1$ and $C_2$ are of the same order, and they both decay as $|i-j|^{-2}$ for $\pi|i-j|/N\ll 1$. For long distances, where $|i-j|\approx N/2$, both $C_1$ and $C_2$ approach zero. The fact that the interaction strengths decay fast with distance suggest that it may be possible to truncate the nonlocal Hamiltonian to a local Hamiltonian without significantly altering the low-energy physics of the model.

\subsection{Local models}
Replacing the exact Hamiltonian with a local one is an advantage experimentally, both because it removes the need to engineer couplings
between distant sites and because it reduces the number of terms in the Hamiltonian. Ultracold atoms in optical lattices provide an
interesting framework for simulating quantum physics. By trapping ultracold atoms in a 1D optical lattice, one naturally gets a model
with hopping between NN sites and on-site interactions \cite{Jaksch}. The hardcore constraint can be implemented by ensuring that the
on-site interaction is large enough. Density-density interactions between neighboring sites can be achieved through dipole-dipole
interactions, see e.g.\ \cite{Trefzger,Sowinski,Baier}.
\begin{figure}
\begin{indented}\item[]
\includegraphics[width=0.6\columnwidth]{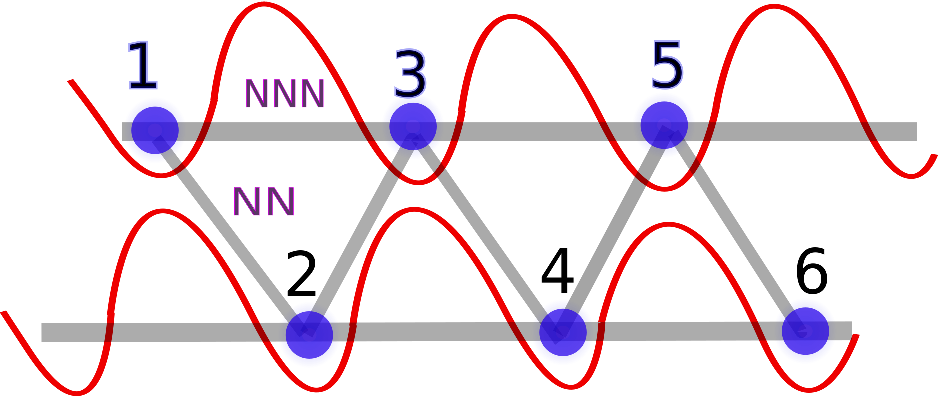}
\caption{Schematic diagram showing a zigzag lattice. The geometry allows for adjusting the strengths of the NNN terms relative
to the strengths of the NN terms.}
\label{zigzag}
\end{indented}
\end{figure}

If one changes the geometry of the 1D lattice into a zigzag lattice, one can have couplings between both NN and NNN sites, as schematically
depicted in Fig.\ \ref{zigzag}. There are already more methods available to realize zigzag lattices with
cold atoms, see e.g.\ \cite{Greschner,Dhar,Zhang,Anisimovas}. Anisimovas {\it et.\ al.\ }have proposed a scheme for realizing a zigzag
lattice \cite{Anisimovas} using ultracold bosons. In their study, the two legs of the ladder correspond to a synthetic dimension given by
two spin states of the atoms, and the tunneling between them can be realized by a laser-assisted process. Subsequently, by employing a
spin-dependent optical lattice with the site position depending on the internal atomic state, a zigzag ladder can be achieved. In another
experimental proposal, Zhang {\it et.\ al.\ }have described a feasible method to achieve the zigzag lattice in one dimension \cite{Zhang}.
They use a superlattice generated by commensurate wavelengths of light beams to realize tunable geometries including zigzag and sawtooth
configurations. Also, in a theoretical study, Greschner {\it et.\ al.\ }have suggested that the zigzag chain may be formed by an incoherent
superposition between a triangular lattice and a 1D supperlattice \cite{Greschner}.

In the following, we will therefore restrict ourselves to models with at most NNN couplings in the Hamiltonian. Specifically, we study
four different models. The first one is the NN model defined by the Hamiltonian
\begin{eqnarray}
H_{\textrm{NN}} = \sum_{<i,j>}
        \left[
        (q-2)w_{ij} - w_{ij}^2
        \right]
        d^{\dagger}_i d_j
        -  \frac{1}{2}(q^2 -q)\sum_{<i,j>}w_{ij}^2 n_in_j. \label{Eqn_NN}
\end{eqnarray}
Here, the sum over $i$ and $j$ is restricted to NN terms as denoted by the symbol $<\ldots>$. (We include both, e.g., $(i,j)=(1,2)$ and $(i,j)=(2,1)$ in the sum, which ensures that the Hamiltonian is Hermitian.) For $q=2$, $H_{\textrm{NN}}$ reduces to the antiferromagnetic spin-$1/2$ $XXX$ Heisenberg Hamiltonian except for an additive constant (this can be seen by applying the transformation in (\ref{trans})).

The second model is the model obtained by truncating the full Hamiltonian to only include terms up to NNN distance, i.e.,
\begin{eqnarray}\label{Eqn_NNN}
H_{\textrm{NNN}} = \sum_{<<i,j>>}
        \left[
        (q-2)w_{ij} - w_{ij}^2
        \right]
        d^{\dagger}_i d_j
        -  \frac{1}{2}(q^2 -q)
        \sum_{<<i,j>>}w_{ij}^2 n_in_j,
\end{eqnarray}
where $<<\ldots>>$ is the sum over NN and NNN terms.

\begin{table}
\caption{\label{Uopt}Optimal values of $U_{1}$ and $U_{2}$.}
\begin{indented}
\lineup
\item[]\begin{tabular}{lcc}
\br
$q$     & $U_{1}$ & $U_{2}$    \\
\mr
\vspace{1.3mm}
2      & 1.00    & 1.00    \\
\vspace{1.3mm}
3      & 1.86    & 0.68       \\
\vspace{1.3mm}
4      & 5.36    & 0.60 \\
\br
\end{tabular}
\end{indented}
\end{table}
\begin{figure}
\begin{indented}\item[]
\includegraphics[width=13.0cm,clip=true, angle = 360]{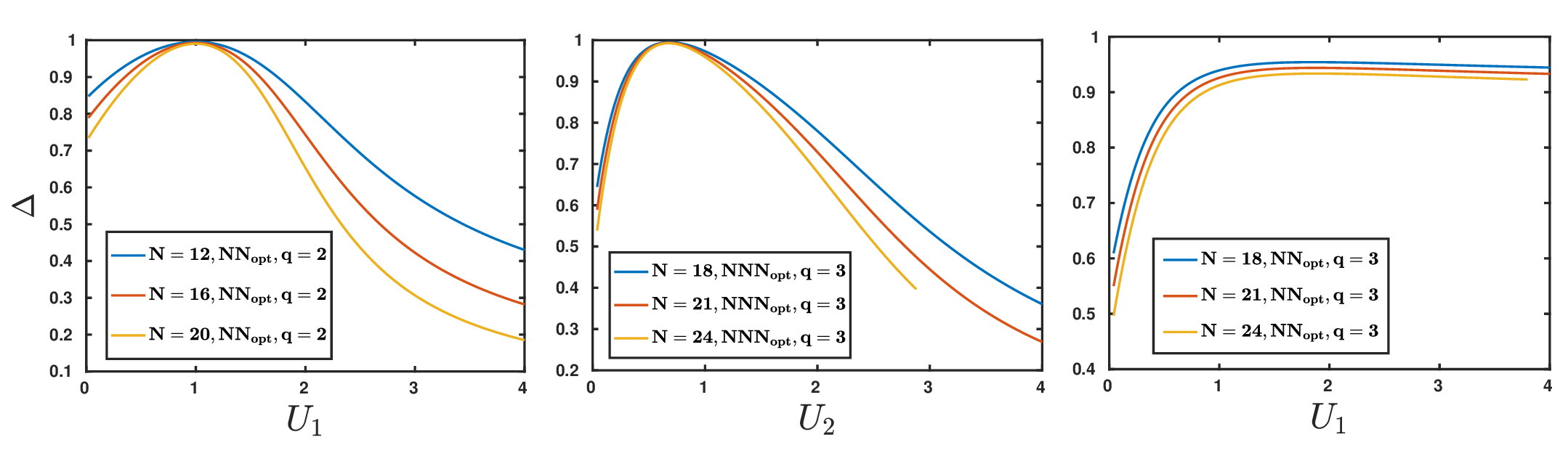}
\includegraphics[width=12.0cm,clip=true, angle = 360]{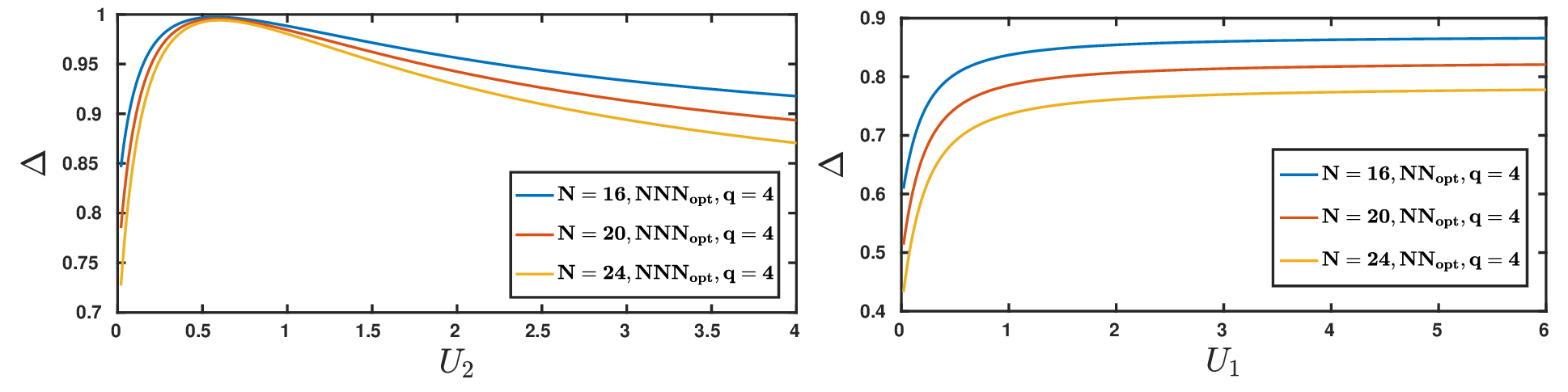}
\caption{Overlap value as a function of optimizing parameters $U_1$ and $U_2$ for $q = 3$, $q = 2$, and $q = 4$.}
\label{FigUopt}
\end{indented}
\end{figure}
Instead of simply cutting the Hamiltonian as done above, one could try to partially compensate for the removed terms by adjusting the
relative strengths of the couplings. We therefore also consider the optimized NN model
\begin{equation}\label{Eqn_Opt_NN}
 H^{\textrm{opt}}_{\textrm{NN}} =\sum_{<i,j>}
        \left[
        (q-2)w_{ij} - w_{ij}^2
        \right]
        d^{\dagger}_i d_j
        -\frac{U_{1}}{2}(q^2 -q)
        \sum_{<i,j>}w_{ij}^2 n_in_j,
\end{equation}
and the optimized NNN model
\begin{equation}\label{Eqn_Opt_NNN}
 H^{\textrm{opt}}_{\textrm{NNN}} =\sum_{<<i,j>>}
        \left[
        (q-2)w_{ij} - w_{ij}^2
        \right]
        d^{\dagger}_i d_j
        -\frac{U_{2}}{2}(q^2 -q)
        \sum_{<<i,j>>}w_{ij}^2 n_in_j.
\end{equation}
In these models, $U_{1}$ and $U_{2}$ are parameters that are chosen to maximize the overlap between
the ground state of $H^{\textrm{opt}}_{\textrm{NN}}$ or $H^{\textrm{opt}}_{\textrm{NNN}}$ with the analytical
state $|\psi_\textrm{Exact}\rangle$ in (\ref{Ana_eqn}). 

We investigate how the overlap varies with $U_1$ and $U_2$ in figure \ref{FigUopt}. For a given $q$, the overlap value is seen to be maximized at a unique value of $U_1$ and $U_2$, independent of the system size $N$. The optimal values are given in table \ref{Uopt}. We do not know why this is the case. For $q=2$, the optimal value is unity, and the optimized models hence coincide with the models that are not optimized. It is interesting to note that the Hamiltonians for $q=2$ and $U_1=U_2=1$ are SU(2) invariant, as all terms can be written in terms of $\vec{S}_j\cdot\vec{S}_k$ (see (\ref{trans})). For $q=2$, the wavefunction (\ref{Ana_eqn}) is also SU(2) invariant. It is hence natural that $U_1=U_2=1$ is the optimal choice.

\section{Properties of the ground states of the local models}\label{sec:gsprop}
We now compute properties of the ground states of the local models to quantify how well the local models reproduce the physics of the exact
models. The results are obtained using exact diagonalization, and we hence consider systems with $q=2$, $3$, and $4$ with at most $32$ lattice
sites and $N/q$ particles. We first show that overlaps per site higher than $0.999$ can be obtained for all these cases with at most optimized NNN
interactions. We then show that the behavior of important properties like entanglement entropy and two-site correlation functions is also
reproduced well (possibly except for a bit of discrepancy for $q=3$ and $N$ even).

\subsection{Overlap between ground state wave functions}
We have calculated the overlap $\Delta$ between the ground state $|\Psi_{\textrm{Local}}\rangle$ of the local
Hamiltonian (\ref{Eqn_NN}), (\ref{Eqn_NNN}), (\ref{Eqn_Opt_NN}) or (\ref{Eqn_Opt_NNN}) and the analytical
wavefunction $|\Psi_{\textrm{Exact}}\rangle$ in (\ref{Ana_eqn}). For a non-degenerate ground state, we define the overlap as
\begin{eqnarray}
 \Delta = |\langle \Psi_{\textrm{Local}} |\Psi_{\textrm{Exact}} \rangle|^2.
\end{eqnarray}
For large enough systems, the overlap is expected to decay exponentially with system size, even if the states deviate only a bit from each other, since the dimension of the Hilbert space increases exponentially in $N$. We therefore also consider the overlap per site $\Delta^{1/N}$, which
is expected to remain roughly constant with system size for large enough $N$. We find that the ground state is non-degenerate for all the system sizes we consider.

The overlap for the fermionic case ($q=3$) is given in table \ref{ov_q3} in the appendix. For a complete analysis, these $\Delta$ values are explicitly calculated with and without optimization for different system sizes. The overlaps are higher for the NNN Hamiltonian than for the NN Hamiltonian, and optimization also improves the overlaps. The overlap decreases with system size, but the overlap per site is almost constant. The overlap per site is already around $0.996$ for only NN interactions without optimization, and for the NNN Hamiltonian with optimzation it is around $0.9997$.

Next, we report the overlap values for half-filling ($q=2$) and quarter-filling ($q = 4$) of the lattice with hardcore bosons. These results are shown in tables \ref{ov_q2} and \ref{ov_q4} in the appendix, respectively. For $q=2$, the NN model is enough to obtain overlaps per site around 0.9995, and these overlaps are slightly improved by considering the NNN model.

The scenario is different for $q=4$. In this case, the overlap per site is around 0.987 for the NN model. Adding NNN interactions
increases the overlap per site to around 0.999. The optimized NN model produces lower overlaps than the NNN model. Optimizing the
NNN model gives overlaps slightly better than for the NNN model. For $q=4$, the average distance between particles is four sites, which
could be part of the explanation why NNN interactions are more important in this case than for $q=2$.

From the above analysis of the overlap, we conclude that for the $q=2$ case, it is sufficient to consider the local Hamiltonian with only
NN interactions to study the ground state properties, whereas for the $q=3$ and $q=4$, the local NNN Hamiltonian with optimization is
an appropriate choice.

\subsection{Entanglement Entropy}

\begin{figure}
\begin{indented}
\item[]
\includegraphics[width=12cm,clip=true, angle = 360]{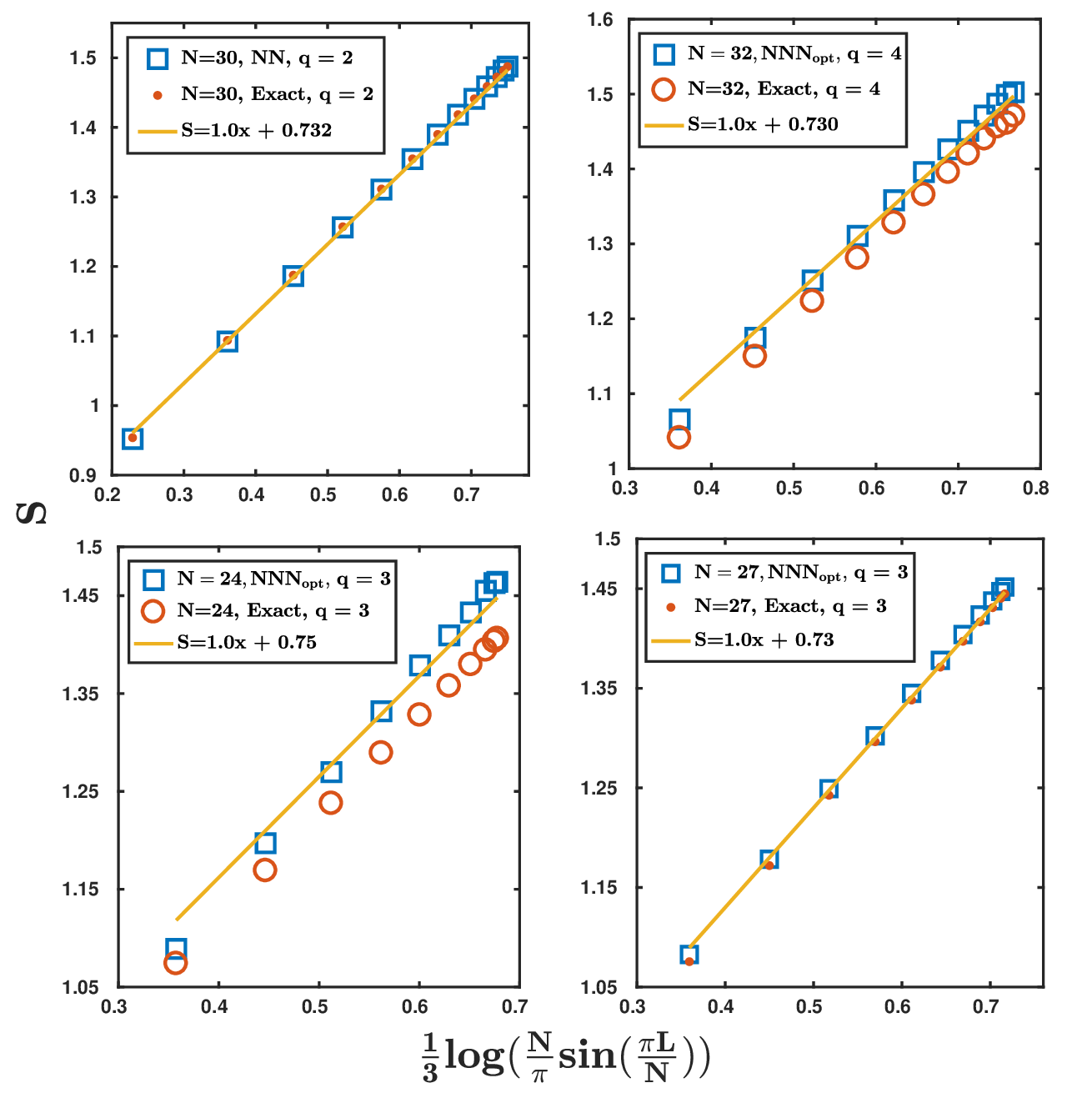}
\caption{Entanglement entropy $S$ as a function of subsystem size $L$ for $q=2$, $q=4$ and $q=3$. The numerical results for $S$ are
computed using the ground states of $H_{\textrm{NN}}$ for $q=2$ and $H_{\textrm{NNN}}^{\textrm{opt}}$ for $q=3$ and $q=4$.
The different plots are for different $q$ values as specified in the legends. The equations for the straight lines with slope $1$ are also shown in the legends.}\label{entropy}
\end{indented}
\end{figure}

We next consider the entanglement entropy. The entanglement entropy quantifies bipartite entanglement between two subsystems and is a typical
measure of entanglement \cite{Bennett}. Besides its own fundamental theoretical interest, a principal reason for the success of the entanglement
entropy as an entanglement measure in extended quantum systems is its universal scaling at 1D conformal critical points. It has already been
investigated that the entanglement entropy and its spectra carry signatures of the underlying conformal field theory: In 1D, the entanglement
entropy of a conformal field theory exhibits a logarithmic violation of the area law and this characteristic has also been observed in lattice
systems \cite{Vidal, Calabrese}.

For a general 1D system in a state $|\Psi \rangle$, one can define the entanglement entropy by partitioning the whole system into two subsystems
$A$ and $B$. The reduced density operator of subsystem $A$ ($B$) is
\begin{eqnarray}
 \rho_A = \mathrm{Tr}_B(\rho), \qquad
 \rho_B = \mathrm{Tr}_A(\rho),
\end{eqnarray}
where $\rho = |\Psi \rangle \langle \Psi|$ is the density operator of the full system. The von Neumann entanglement entropy of subsystem $A$ is
then defined as
\begin{eqnarray}
 S(A) = -\mathrm{Tr}_A[\rho_A \ln(\rho_A)].
\end{eqnarray}
It can be shown that $S\equiv S(A) = S(B)$, so that only the partitioning of the system is important.

One can also express the von Neumann entanglement entropy in terms of the eigenvalues $\lambda_{A,i}$ of $\rho_A$ as
\begin{eqnarray}
 S = -\sum_{i} \lambda_{A,i} \ln(\lambda_{A,i}).
\end{eqnarray}
We use this formula to calculate the von Neumann entanglement entropy for our 1D system. We consider subsystems that consist of $L$ consecutive
sites. Due to the translational invariance of the models we are considering, all such entanglement entropies can be expressed in terms of the
entanglement entropies obtained when region $A$ is site number $1$ to site number $L$ with $L\in\{1,2,\ldots,N/2\}$ for $N$ even
and $L\in\{1,2,\ldots,(N+1)/2\}$ for $N$ odd.

The von Neumann entanglement entropy of a critical system in 1D is given by \cite{ee}
\begin{equation}
S=\frac{c}{3}\log\left[\frac{N}{\pi}\sin\left(\frac{\pi L}{N}\right)\right]+\textrm{constant} \label{EE},
\end{equation}
when $L$ is large compared to $1$. The central charge $c$ was found to be $c=1$ for the exact model in \cite{TuNJP}. Plotting $S$ as a function of $(1/3)\log[(N/\pi)\sin(\pi L/N)]$, we should hence compare to a line of slope $1$.

Plots of the entanglement entropy for $q=3$, $2$, and $4$ and different system sizes are shown in Fig.\ \ref{entropy}. For $q=2$, the data for the NN model and the exact model are seen to fit very well and also fit well with a line of slope $1$. For $q=3$, there is a good match for $N$ odd, while there is a bit of deviation for $N$ even. For $q=4$, the data for the local model are displaced by a constant, but the slope for $L$ not too small fits well with $1$. The central charge is hence the same as in the exact model.

\subsection{Correlation function}
\begin{figure}
\includegraphics[width=50mm,clip=true, angle = 360]{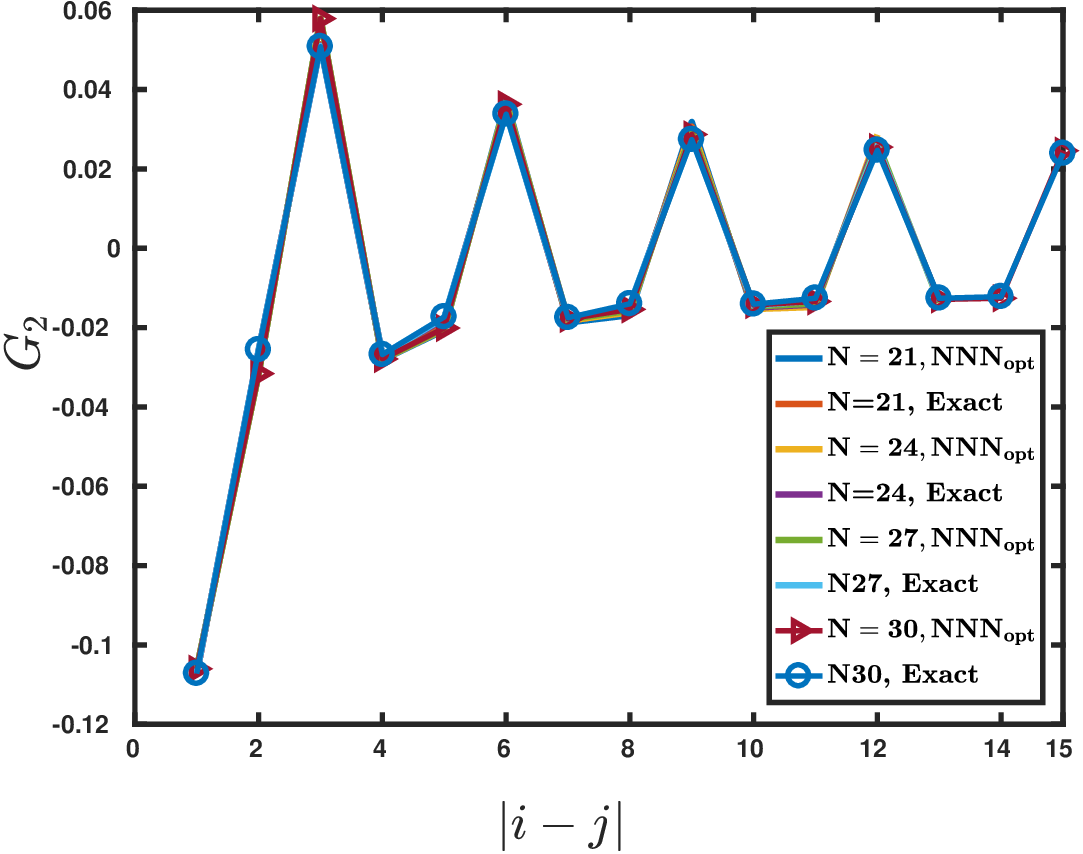}\hfill
\includegraphics[width=52mm,clip=true, angle = 360]{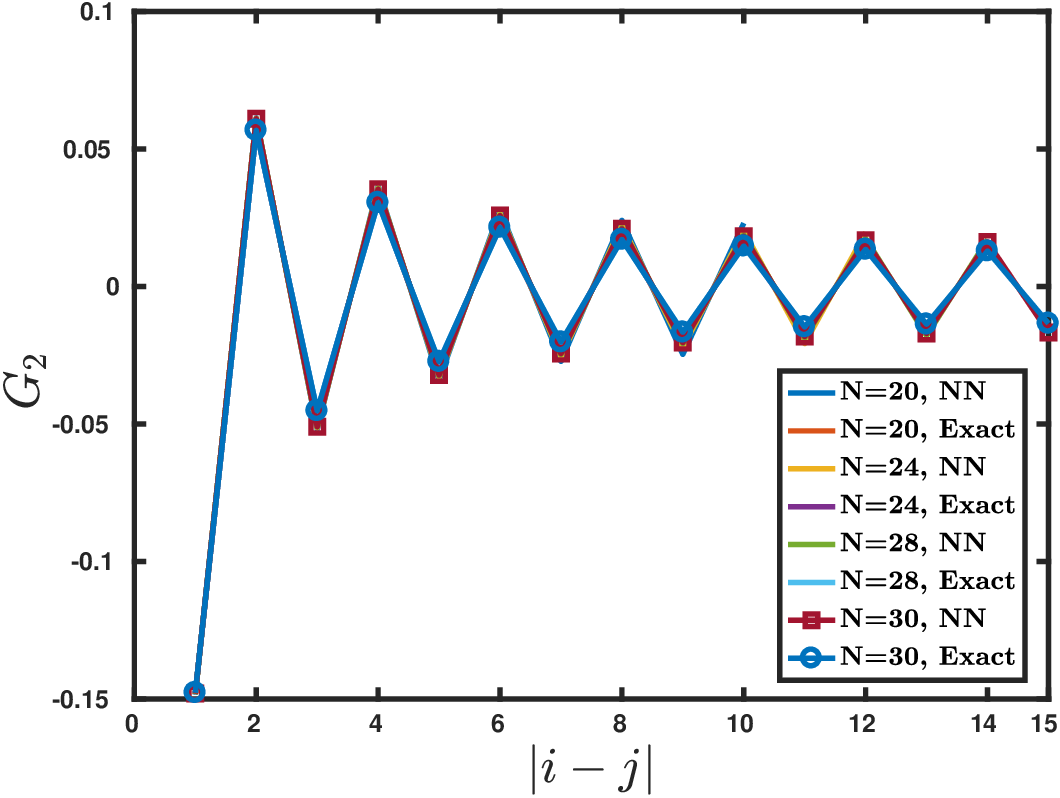}\hfill
\includegraphics[width=46mm,clip=true, angle = 360]{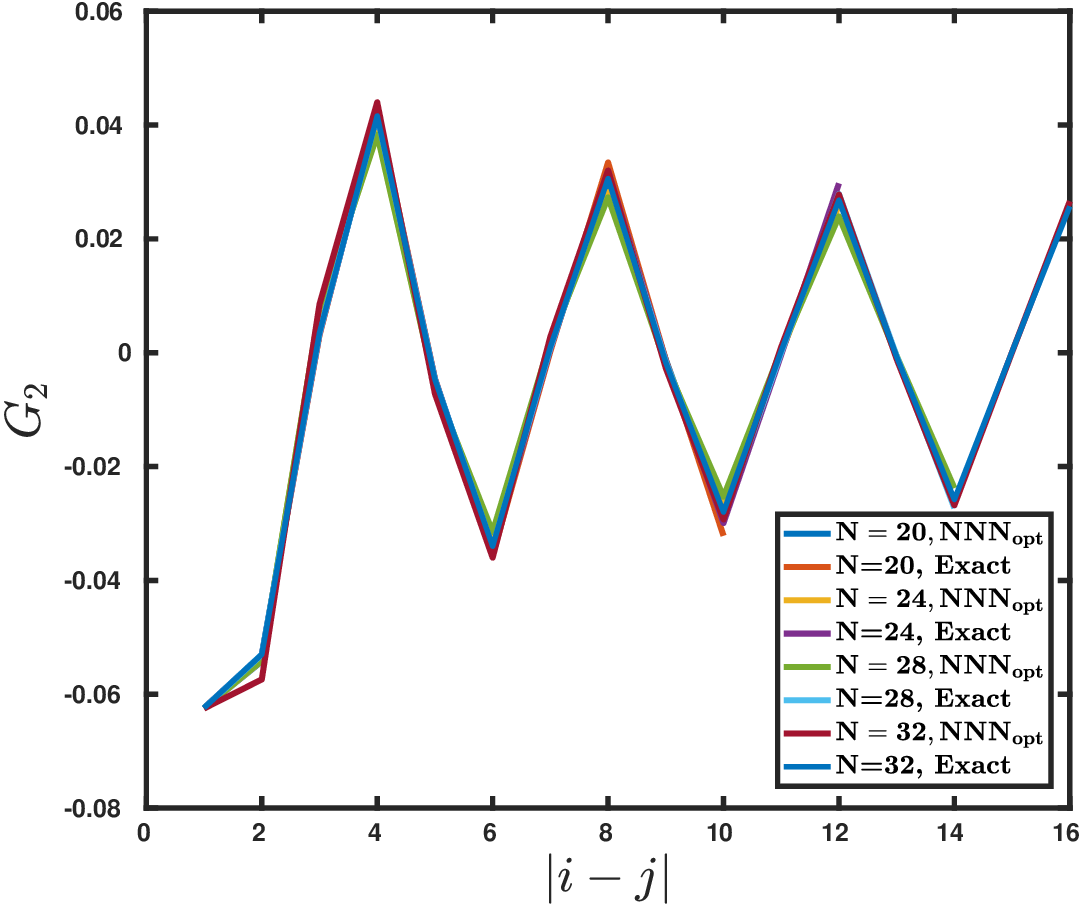}
\caption{Correlation function $G_2$ for $q=3$ (left), $q=2$ (middle), and $q=4$ (right). Results are shown both for the wavefunction $|\Psi_{\textrm{Exact}}\rangle$ and for the ground state of the local Hamiltonian ($H_{\textrm{NNN}}^{\textrm{opt}}$ for $q=3$ and $q=4$ and $H_{\textrm{NN}}$ for $q=2$).}
\label{G2}
\end{figure}

Correlation functions are also an important characteristic property for determining the physics of a system. Here, we compute the two-site correlation
function defined as
\begin{eqnarray}
G_2 &=& \langle \Psi |n_i n_j | \Psi \rangle - \langle \Psi | n_i | \Psi \rangle \langle \Psi | n_j | \Psi \rangle,
\end{eqnarray}
where $|\Psi\rangle$ is the state of the system. In our case, $\langle \Psi | n_j | \Psi \rangle = 1/q$ for all $j$. In the exact model, $G_2$ decays
with the distance between the points to the power $-2/q$ and also shows oscillations with period $q$. For the local models, it is difficult to determine
the precise behavior of the decay, since we cannot go to sufficiently large system sizes. Instead we simply compare the values of $G_2$ for the local and
the exact models.

Plots of $G_2$ are shown for $q=3$, $2$, and $4$ and different system sizes in Fig.\ \ref{G2}. In all cases, there is a good agreement between the local and the exact models.

\section{Low-lying excited states}\label{sec:esprop}

\begin{figure}
\begin{indented}\item[]
\includegraphics[width=13.2cm,clip=true, angle = 360]{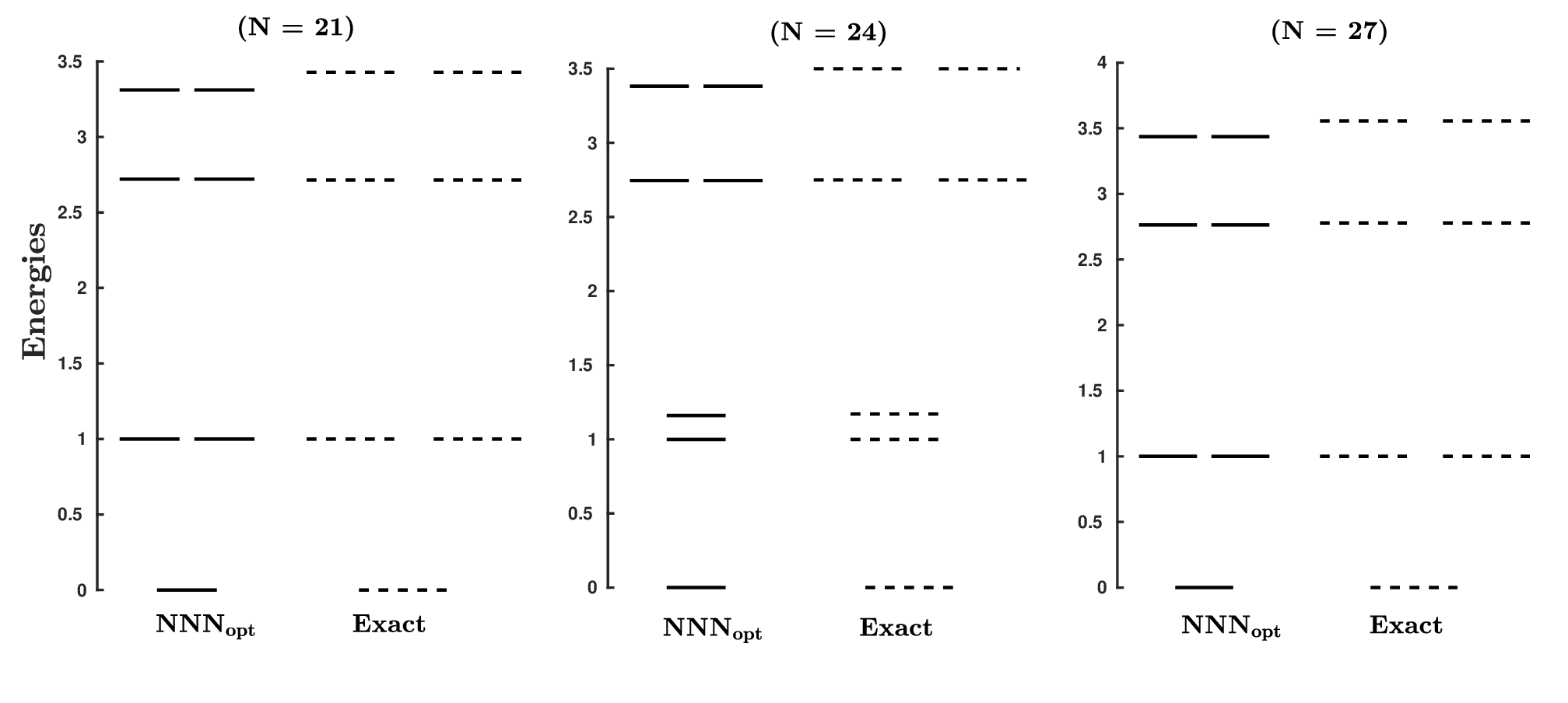}
\caption{Energies of the $7$ lowest states of the NNN$_{\textrm{opt}}$ model (\ref{Eqn_Opt_NNN}) (shown as solid lines) and of the exact model (\ref{Ham}) (shown as dashed lines) for $q = 3$ (the ground state energy has been set to zero, and the energies have been scaled by a constant factor such that the first excited state is at energy 1). The different plots are for $21$, $24$, and $27$ sites, respectively, as indicated.}
\label{Engq3}
\end{indented}
\end{figure}

\begin{figure}
\begin{indented}\item[]
\includegraphics[width=13.2cm,clip=true, angle = 360]{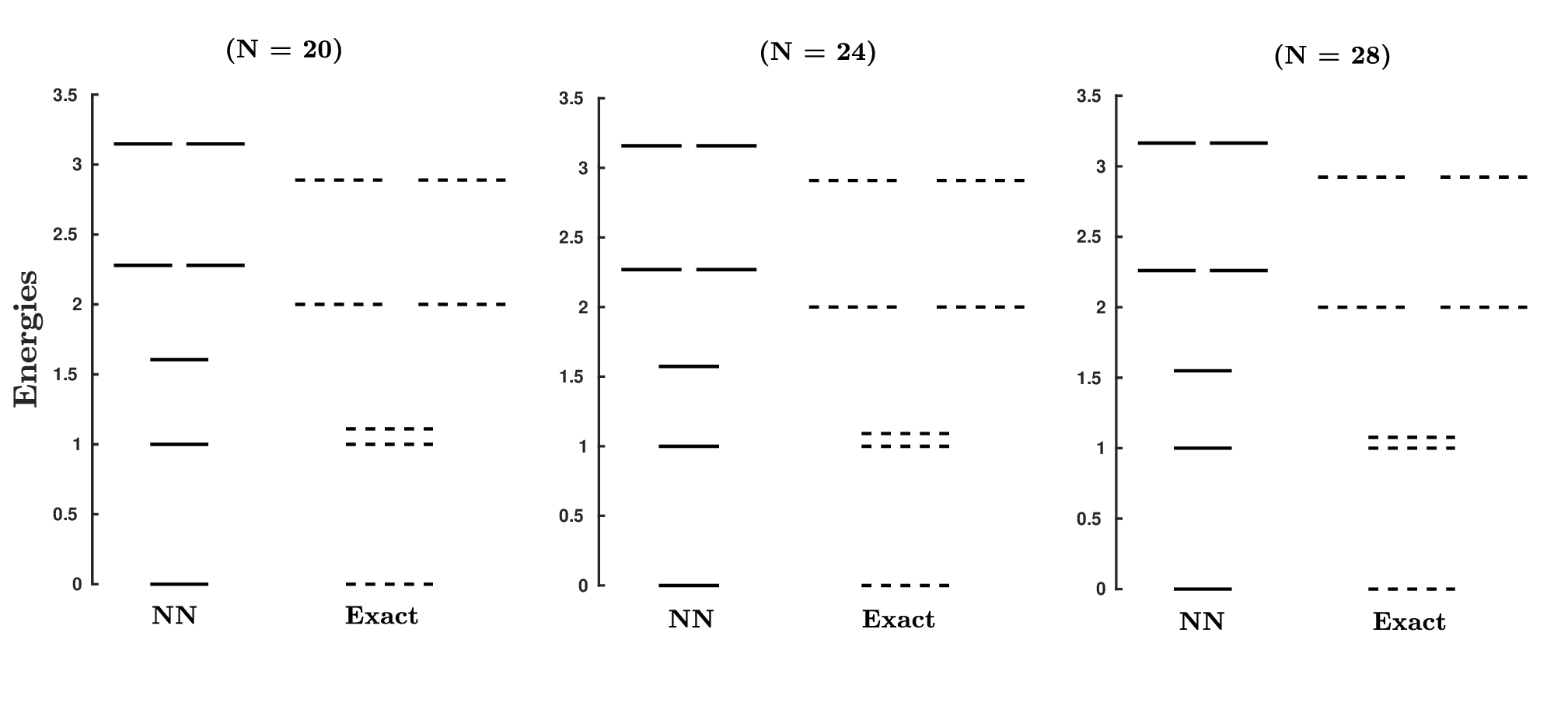}
\caption{Energies of the $7$ lowest states of the NN model (\ref{Eqn_NN}) (shown as solid lines) and of the exact model (\ref{Ham}) (shown as dashed lines) for $q = 2$ (the ground state energy has been set to zero, and the energies have been scaled by a constant factor such that the first excited state is at energy 1). The different plots are for $20$, $24$, and $28$ sites, respectively, as indicated.}
\label{Engq2}
\end{indented}
\end{figure}

\begin{figure}
\begin{indented}\item[]
\includegraphics[width=13.2cm,clip=true, angle = 360]{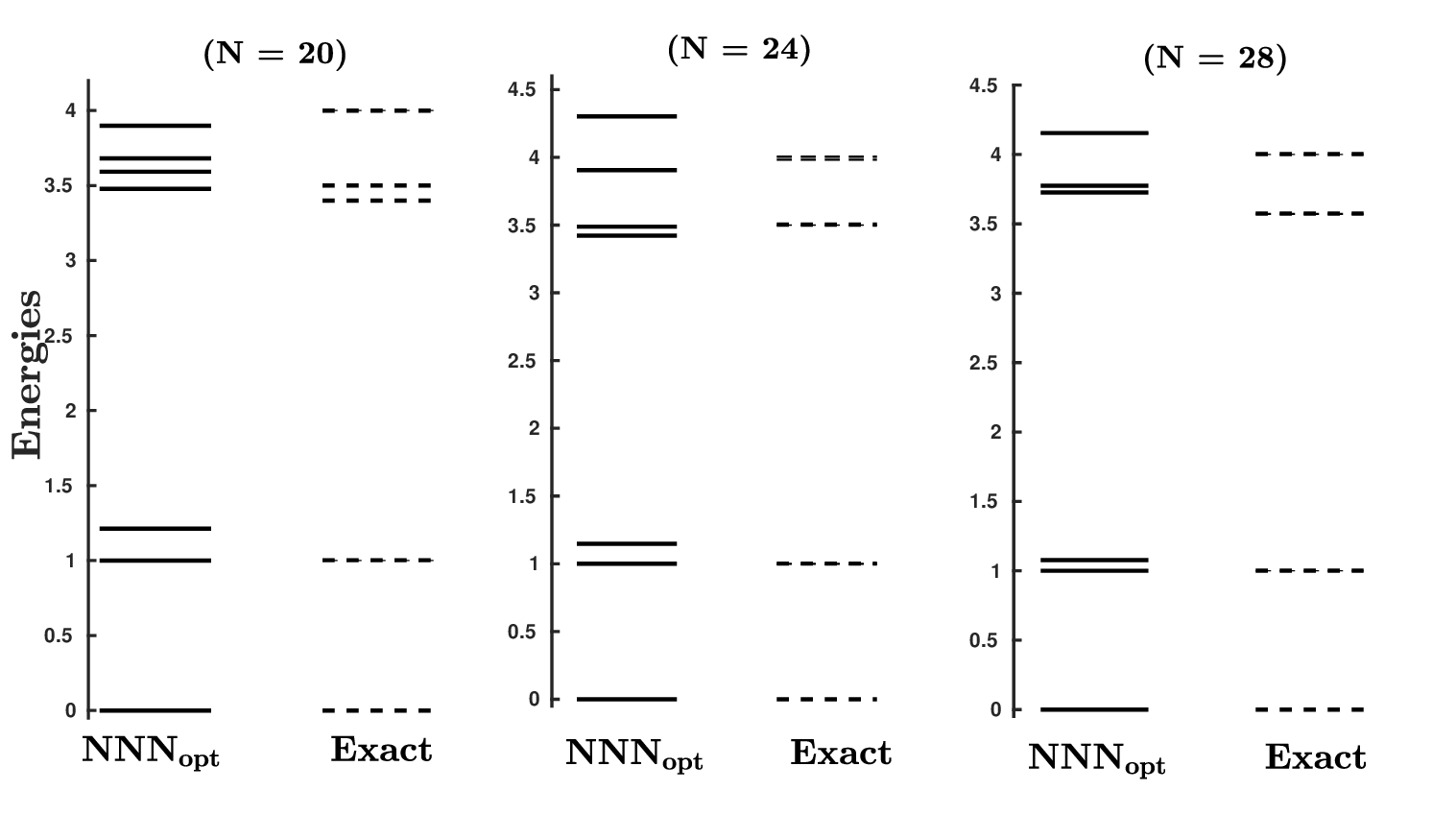}
\caption{Energies of the $7$ lowest states of the NNN$_\textrm{opt}$ model (\ref{Eqn_Opt_NNN}) (shown as solid lines) and of the exact model (\ref{Ham}) (shown as dashed lines) for $q = 4$ (the ground state energy has been set to zero, and the energies have been scaled by a constant factor such that the first excited state is at energy 1). The different plots are for $20$, $24$, and
$28$ sites, respectively, as indicated.}
\label{Engq4}
\end{indented}
\end{figure}

Achieving a good overlap for the ground state wave functions is a good starting point, but if the system is at nonzero temperature or we are interested in dynamics, it is also important that the low-lying excited states are not affected significantly in going from the exact to the local Hamiltonian, and we now investigate this question.

The energies of the seven lowest states are plotted in Figs.\ \ref{Engq3}, \ref{Engq2}, and \ref{Engq4} for $q=3$, $2$, and $4$, respectively, for both a local model and the exact model. We have added a constant to the energies and divided by a constant factor to fix the ground state energy to zero and the energy of the first excited state to one. For $q=3$, it is seen that the energies of the lowest excited states of the local model fit accurately with the energies of the corresponding states for the exact model, in particular for the first four excited states. For $q=2$, the energies of the lowest excited states in the local model are generally shifted compared to the exact model, but the degeneracies remain the same. In particular, it is seen that the energy difference between the first and the second excited state is much larger for the local model than for the exact model. For $q=4$, it is again the case that the energies of the lowest excited states in the local model are shifted compared to the exact model, but the degeneracies remain the same. It is interesting to note that the plots for a given value of $q$ are almost independent of $N$ for $q=2$ and $q=3$, but not for $q=4$. This shows that finite size effects play a role for $q=4$. Our exact diagonalization codes do, however, not allow us to go to large enough systems to eliminate these effects.

We next compute overlaps to investigate how well the wavefunctions of the low-energy excited states of the local models match with the corresponding eigenstates of the exact models. We find that the overlap per site is higher than $0.998$ in all the considered cases, which shows that there is an excellent agreement. Details of the results can be found in tables \ref{ex_ovq3}, \ref{ex_ovq2}, and \ref{ex_ovq4} in the appendix. The fact that the overlaps are nonzero also shows that there is agreement between all quantum numbers of the states of the local Hamiltonian and the states of the exact Hamiltonian.

The conclusion of this section is that the low-energy excited state wavefunctions of the local models are close to the corresponding wavefunctions of the exact models, but the energies may be shifted. If we prepare the system in a given low-energy excited state, we hence expect the physics to be practically the same for the local and the exact model. If we instead consider a superposition of different excited states, the relative phases between the terms in the superposition may have different time variation.

\section{Conclusion}\label{sec:conclusion}
We have started from a class of one-dimensional models derived from conformal field theory and used exact diagonalization to investigate to what extent the same physics is realized by different models containing only NN and possibly NNN couplings. For the $q=2$ model we found that already the NN model, which is equivalent to the antiferromagnetic $XXX$ Heisenberg model, gives a good description. For $q=3$ and $q=4$, an optimized model with both NN and NNN couplings is preferred. With these models we find that the overlap per site between the ground states of the local models and the ground states of the exact models is larger than $0.999$ for the considered system sizes. The overlaps per site for the low-lying excited state are larger than $0.998$. We also find that the ground state entanglement entropy and the correlation functions are reproduced well by the local models, except that there is a bit of discrepancy for $q=3$ for $N$ even for the entanglement entropy. The low-energy spectra are modified to some extent in the local models compared to the exact models. For $q=4$, the modification varies with system size, so finite size effects play a role for the considered system sizes.

The high agreement between the ground state properties and the excited states suggest that most of the physics will remain the same in the local
models. This paves the way to experimentally realize the physics in optical lattices using ultracold atoms, since the local models are simpler to
realize experimentally than the corresponding exact models derived from conformal field theory. The ingredients needed to realize the different
terms in the Hamiltonian are already available experimentally. The observation that the local models resemble the exact models also shows that
conformal field theory can not just be used to obtain exact, nonlocal models, but is also a powerful tool to obtain interesting, local models.

\section*{Acknowledgments}
This work has in part been supported by the Villum Foundation. DKN would like to thank the Max Planck Institute for the Physics of Complex Systems for hospitality during visits to the institute.

\appendix
\section{Ground state overlap}\label{sec:gsov}
In tables \ref{ov_q3}, \ref{ov_q2}, and \ref{ov_q4}, we present the complete data for the ground state overlap for different $q$ and $N$ values.

\begin{table}
\caption{Overlap $\Delta$ and overlap per site $\Delta^{1/N}$ between the state $|\psi_\textrm{Exact}\rangle$ in (\ref{Ana_eqn}) and the
ground state of $H_{\textrm{NN}}$, $H_{\textrm{NNN}}$, $H_{\textrm{NN}}^{\textrm{opt}}$, or $H_{\textrm{NNN}}^{\textrm{opt}}$ for $q=3$.
The different rows are for different numbers of lattice sites $N$.}
\vspace{2.5mm}
\begin{indented}
\lineup
\item[]\begin{tabular}{lccccccccccc}
\hline
\hline \\
\vspace{1.0mm}
$N$   &  &\multicolumn{4}{c}{$\Delta$}& &\multicolumn{4}{c}{$\Delta^{1/N}$} \\
\cline{3-6} \cline{8-11} \\
      &  & NN & NNN & NN$_{\textrm{opt}}$ & NNN$_{\textrm{opt}}$ & & NN & NNN & NN$_{\textrm{opt}}$ & NNN$_{\textrm{opt}}$ \\
\hline \\
\vspace{2.0mm}
15   & & 0.953  & 0.981 &0.965  &0.996 & &0.9968  &0.9987 & 0.9976  &0.9998 \\
\vspace{2.3mm}
18   & & 0.939  & 0.973 &0.954  &0.995 & &0.9965  &0.9985 & 0.9974  &0.9997 \\
\vspace{2.3mm}
21   & & 0.926  & 0.966 &0.944  &0.994 & &0.9963  &0.9984 & 0.9973  &0.9997 \\
\vspace{2.3mm}
24   & & 0.913  & 0.959 &0.934  &0.992 & &0.9962  &0.9983 & 0.9971  &0.9997 \\
\vspace{2.3mm}
27   & & 0.899  & 0.953 &0.923  &0.991 & &0.9961  &0.9982 & 0.9971  &0.9997 \\
\vspace{2.3mm}
30   & & 0.880  & 0.944 &0.912  &0.987 & &0.9957  &0.9981 & 0.9970  &0.9996 \\
\hline
\hline
\end{tabular}
\end{indented}
\label{ov_q3}
\end{table}

\begin{table}
\caption{Overlap $\Delta$ and overlap per site $\Delta^{1/N}$ between the state $|\psi_\textrm{Exact}\rangle$ in (\ref{Ana_eqn}) and
the ground state of $H_{\textrm{NN}}$ or $H_{\textrm{NNN}}$ for $q=2$. The different rows are for different numbers of lattice sites $N$.}
\vspace{2.5mm}
\begin{indented}
\lineup
\item[]\begin{tabular}{lcccccc}
\hline
\hline \\
$N$      & \multicolumn{2}{c}{$\Delta$} & & \multicolumn{2}{c}{$\Delta^{1/N}$}   \\
 \cline{2-3}\cline{5-6}\\
                  & NN  & NNN & & NN & NNN  \\
\hline \\
\vspace{2.3mm}
16    & 0.9930    &0.9960  &  &0.9996   &0.9997   \\
\vspace{2.3mm}
18    & 0.9917    &0.9950  &  &0.9995   &0.9997   \\
\vspace{2.3mm}
20    & 0.9904    &0.9940  &  &0.9995   &0.9997   \\
\vspace{2.3mm}
22    & 0.9891    &0.9931  &  &0.9995   &0.9997   \\
\vspace{2.3mm}
24    & 0.9878    &0.9922  &  &0.9995   &0.9997   \\
\vspace{2.3mm}
26    & 0.9864    &0.9912  &  &0.9995   &0.9997   \\
\vspace{2.3mm}
28    & 0.9852    &0.9903  &  &0.9995   &0.9997   \\
\vspace{2.3mm}
30    & 0.9839    &0.9894  &  &0.9995   &0.9996   \\
\hline
\hline
\end{tabular}
\end{indented}
\label{ov_q2}
\end{table}

\begin{table}
\caption{Overlap $\Delta$ and overlap per site $\Delta^{1/N}$ between the state $|\psi_\textrm{Exact}\rangle$ in (\ref{Ana_eqn})
and the ground state of $H_{\textrm{NN}}$, $H_{\textrm{NNN}}$, $H_{\textrm{NN}}^{\textrm{opt}}$, or $H_{\textrm{NNN}}^{\textrm{opt}}$
for $q=4$. The different rows are for different numbers of lattice sites $N$.}
\vspace{2.5mm}
\begin{indented}
\lineup
\item[] \begin{tabular}{lccccccccccc}
\br
$N$  & & \multicolumn{4}{c}{$\Delta$} & \multicolumn{4}{c}{$\Delta^{1/N}$}   \\
 \cline{3-6}\cline{8-11}\\
     & & NN  & NNN & NN$_{\textrm{opt}}$  & NNN$_{\textrm{opt}}$ & & NN & NNN & NN$_{\textrm{opt}}$ & NNN$_{\textrm{opt}}$ \\
\hline \\
\vspace{2.3mm}
16   & &0.837   &0.988 &0.865 &0.997 & &0.9881  &0.9993 &0.9910 &0.9998 \\
\vspace{2.3mm}
20   & &0.785   &0.984 &0.820 &0.996 & &0.9880  &0.9992 &0.9901 &0.9998 \\
\vspace{2.3mm}
24   & &0.736   &0.980 &0.777 &0.994 & &0.9873  &0.9992 &0.9895 &0.9998 \\
\vspace{2.3mm}
28   & &0.690   &0.976 &0.736 &0.992 & &0.9868  &0.9992 &0.9891 &0.9997 \\
\vspace{2.3mm}
32   & &0.672   &0.968 &0.702 &0.973 & &0.9877  &0.9990 &0.9890 &0.9991 \\
\br
\end{tabular}
\end{indented}
\label{ov_q4}
\end{table}

\section{Excited state overlap}\label{sec:exov}
In tables \ref{ex_ovq3}, \ref{ex_ovq2}, and \ref{ex_ovq4}, we present complete data for the overlap of the low-lying excited states considered in our calculation.

\begin{table}
\caption{Overlap $\Delta$ and overlap per site $\Delta^{1/N}$ between the first six excited states of the NNN$_{\textrm{opt}}$ Hamiltonian and
the first six excited states of the nonlocal Hamiltonian (\ref{Ham}) for $q=3$. The number of sites is $N = 21$, $24$, and $27$, respectively.}
\vspace{2.5mm}
\begin{indented}
\lineup
\item[] \begin{tabular}{lcccccccc}
\hline
\hline \\
State  & \multicolumn{3}{c}{$\Delta$} & & \multicolumn{3}{c}{$\Delta^{1/N}$}   \\
 \cline{2-4}  \cline{6-8}\\
  &$N = 21$ & $N = 24$ & $N =27$ & &$N = 21$ & $N = 24$ & $N =27$  \\
\hline \\
\vspace{2.3mm}
$2$  & 0.9933 &0.9921  & 0.9908  & &0.9997  &0.9997  &0.9997  \\
\vspace{2.3mm}
$3$  & 0.9933 &0.9921  & 0.9908  & &0.9997  &0.9997  &0.9997  \\
\vspace{2.3mm}
$4$  & 0.9894 &0.9887  & 0.9878  & &0.9995  &0.9995  &0.9995  \\
\vspace{2.3mm}
$5$  & 0.9894 &0.9887  & 0.9878  & &0.9995  &0.9995  &0.9995  \\
\vspace{2.3mm}
$6$  &0.9893  &0.9887  &0.9795   & &0.9995  &0.9995  &0.9992  \\
\vspace{2.3mm}
$7$  &0.9893  &0.9887  &0.9795   & &0.9995  &0.9995  &0.9992  \\
\hline
\hline
\end{tabular}
\end{indented}
\label{ex_ovq3}
\end{table}

\begin{table}
\caption{Overlap $\Delta$ and overlap per site $\Delta^{1/N}$ between the first six excited states of the NN Hamiltonian and the first six
excited states of the nonlocal Hamiltonian (\ref{Ham}) for $q=2$. The number of sites is $N = 20$, $24$, and $28$, respectively.}
\vspace{2.5mm}
\begin{indented}
\lineup
\item[] \begin{tabular}{lccccccc}
\hline
\hline \\
& \multicolumn{3}{c}{$\Delta$}&  & \multicolumn{3}{c}{$\Delta^{1/N}$}  \\
\cline{2-4} \cline{6-8}\\
State  & $N=20$  & $N=24$  & $N=28$ & & $N=20$  & $N=24$  & $N=28$  \\
\hline \\
\vspace{2.3mm}
$2$   &0.9900  &0.9879   &0.9856 & & 0.9995  &0.9995  &0.9995    \\
\vspace{2.3mm}
$3$   &0.9707  &0.9668   &0.9634 & & 0.9985  &0.9986  &0.9987    \\
\vspace{2.3mm}
$4$   &0.9914  &0.9800   &0.9772 & & 0.9996  &0.9992  &0.9992    \\
\vspace{2.3mm}
$5$   &0.9914  &0.9800   &0.9772 & & 0.9996  &0.9992  &0.9992    \\
\vspace{2.3mm}
$6$   &0.9894  &0.9763   &0.9739 & & 0.9995  &0.9990  &0.9991    \\
\vspace{2.3mm}
$7$   &0.9894  &0.9763   &0.9739 & & 0.9995  &0.9990  &0.9991    \\
\hline
\hline
\end{tabular}
\end{indented}
\label{ex_ovq2}
\end{table}

\begin{table}
\caption{Overlap $\Delta$ and overlap per site $\Delta^{1/N}$ between the first six excited states of the NNN$_{\textrm{opt}}$ Hamiltonian and
the first six excited states of the nonlocal Hamiltonian (\ref{Ham}) for $q=4$. The number of sites is $N = 20$, $24$, and $28$, respectively.}
\vspace{2.5mm}
\begin{indented}
\lineup
\item[] \begin{tabular}{lccccccc}
\hline
\hline \\
State  & \multicolumn{3}{c}{$\Delta$} & &\multicolumn{3}{c}{$\Delta^{1/N}$}  \\
\cline{2-4} \cline{6-8}\\
& $N=20$ & $N=24$ & $N=28$ & & $N=20$ & $N=24$ & $N=28$ \\
\hline \\
\vspace{2.3mm}
$2$   &0.9915  &0.9906  &0.9893  &  &0.9996  &0.9996  &0.9996 \\
\vspace{2.3mm}
$3$   &0.9933  &0.9917  &0.9900  &  &0.9997  &0.9997  &0.9996 \\
\vspace{2.3mm}
$4$   &0.9860  &0.9845  &0.9828  &  &0.9993  &0.9993  &0.9994 \\
\vspace{2.3mm}
$5$   &0.9883  &0.9858  &0.9835  &  &0.9994  &0.9994  &0.9994 \\
\vspace{2.3mm}
$6$   &0.9632  &0.9678  &0.9743  &  &0.9981  &0.9986  &0.9991 \\
\vspace{2.3mm}
$7$   &0.9697  &0.9715  &0.9624  &  &0.9985  &0.9988  &0.9986 \\
\hline
\hline
\end{tabular}
\end{indented}
\label{ex_ovq4}
\end{table}

\end{document}